\documentclass{article}
\usepackage{spconf}
\usepackage{diagbox} 
\usepackage{graphicx}
\usepackage{epstopdf}
\usepackage{psfrag}
\usepackage{subfigure}
\usepackage{url}
\usepackage{stfloats}
\usepackage{amsfonts,amssymb,amsmath,bm,paralist,theorem,cite,ifthen,color,nccmath}
\usepackage[font=normalsize,labelfont=bf]{caption}
\usepackage{calc}
\usepackage{enumerate}
\usepackage{multirow}
\usepackage{makecell}
\usepackage[ruled]{algorithm2e}
\usepackage{setspace}
\usepackage{array}
\usepackage{float}
\usepackage{soul}
\usepackage{bbm}
\usepackage{cases}
\usepackage{lipsum} 
\usepackage{etoolbox}
\usepackage[colorlinks,
            linkcolor=blue,
            anchorcolor=red,
            citecolor=red]{hyperref}

\SetKwInOut{Input}{Input}
\SetKwInOut{Output}{Output}

\newcommand{\T}{{\scriptscriptstyle\mathsf{T}}}
\renewcommand{\H}{{\scriptscriptstyle\mathsf{H}}}

\allowdisplaybreaks

\graphicspath{{Figures/}}

\newcommand\Ccl{\ensuremath{\mathcal{C}}}
\newcommand\Dcl{\ensuremath{\mathcal{D}}}

\newcommand\Ncl{\ensuremath{\mathcal{N}}}

\newcommand\Icl{\ensuremath{\mathcal{I}}}

\newcommand\Tcl{\ensuremath{\mathcal{T}}}

\newcommand\Vcl{\ensuremath{\mathcal{V}}}

\newcommand\Xcl{\ensuremath{\mathcal{X}}}
\newcommand\Lcl{\ensuremath{\mathcal{L}}}
\newcommand\Zcl{\ensuremath{\mathcal{Z}}}

\newcommand\Cs{\ensuremath{{\mathbb{C}}}}
\newcommand\Es{\ensuremath{{\mathbb{E}}}}
\newcommand\Rs{\ensuremath{{\mathbb{R}}}}

\newcommand\Xb{\ensuremath{ \mathbf{X} }}

\newcommand\Zb{\ensuremath{ \mathbf{Z} }}

\newcommand\Pbb{\ensuremath{{\mathbb{P}}}}

\newcommand\bb{\ensuremath{ \mathbf{b} }}

\newcommand\hb{\ensuremath{ \mathbf{h} }}

\newcommand\pb{\ensuremath{ \mathbf{p} }}

\newcommand\vb{\ensuremath{ \mathbf{v} }}

\newcommand\zb{\ensuremath{ \mathbf{z} }}

 \usepackage{afterpage}

\usepackage{etoolbox}
\newcommand{\zerodisplayskips}{%
  \setlength{\abovedisplayskip}{4pt}%
  \setlength{\belowdisplayskip}{4pt}%
  \setlength{\abovedisplayshortskip}{4pt}%
  \setlength{\belowdisplayshortskip}{4pt}}
\appto{\normalsize}{\zerodisplayskips}
\appto{\small}{\zerodisplayskips}
\appto{\footnotesize}{\zerodisplayskips}

 \usepackage[all=normal,paragraphs=tight,floats=normal,mathspacing=normal,wordspacing=tight,charwidths=tight,mathdisplays=normal,leading=normal]{savetrees}

\usepackage[font=footnotesize,labelfont=bf]{caption}

\usepackage{enumitem}

\title{\vspace{-12mm}Attention-Enhanced Learning for Sensing-Assisted Long-Term \\ Beam Tracking in mmWave Communications\vspace{-3mm}}
\name{Mengyuan Ma$^*$, Nhan Thanh Nguyen$^*$, Nir Shlezinger$^\dagger$, Yonina C. Eldar$^\S$, and Markku Juntti$^*$
\vspace{-3mm}}
\address{$^*$Centre for Wireless Communications (CWC), University of Oulu, Finland \\
$^\dagger$School of ECE, Ben-Gurion University of the Negev, Beer-Sheva, Israel \\
$^\S$Faculty of Math and CS, Weizmann Institute of Science, Rehovot, Israel\\
\{mengyuan.ma,  nhan.nguyen, markku.juntti\}@oulu.fi; nirshl@bgu.ac.il; yonina.eldar@weizmann.ac.il\vspace{-5mm}}
\begin{document}
%
\maketitle
\begin{abstract}
Beam training and prediction in millimeter-wave communications are highly challenging due to fast time-varying channels and sensitivity to blockages and mobility. In this context, infrastructure-mounted cameras can capture rich environmental information that can facilitate beam tracking design. In this work, we develop an efficient attention-enhanced machine learning model for long-term beam tracking built upon convolutional neural networks and gated recurrent units to predict both current and future beams from past observed images. The integrated temporal attention mechanism substantially improves its predictive performance. Numerical results demonstrate that the proposed design achieves Top-5 beam prediction accuracies exceeding $90\%$ across both current and six future time slots, significantly reducing overhead arising from sensing and processing for beam training. It further attains $97\%$ of state-of-the-art performance with only $3\%$ of the computational complexity.
\end{abstract}

\section{Introduction}\label{sec:intro}

In millimeter wave (mmWave) communications, beam training requires  significant overhead in terms of signaling, latency, and power consumption \cite{imran2024environment}. 
%
%
As a solution, various sensory information have recently been explored for beam prediction, including the user position captured by Global Navigation Satellite System (GNSS) \cite{Rezaie2022deep,morais2023position}, the point cloud scanned by LiDAR \cite{Jiang2024LiDAR,Klautau2019LiDAR}, radar-sensed data \cite{Demirhan2022radar, Luo2023millimeter}, and vision information from cameras \cite{Yang2023Environment,charan2022vision,imran2024environment,Xu20203D}. Such approaches fall under the category of sensing-assisted communications, which is among the main use cases in  Integrated Sensing and Communications (ISAC). 
%

 The joint exploitation of multiple sensing modalities for beam training have been investigated in \cite{charan2022vision,cui2024sensing,Tariq2024deep,tian2023multimodal,shi2024multimodal,park2025resource,zhu2025advancing,zhang2025multimodal}. However, effectively extracting features of different sensing modalities and performing efficient feature fusion typically requires high complexity machine learning (ML) models, posing a challenge for practical implementation. For example, multiple Transformer architectures \cite{vaswani2017attention} were adopted in \cite{cui2024sensing,Tariq2024deep,tian2023multimodal,park2025resource}. 
 Moreover, most existing works on sensing-aided beam management focus on predicting only the current beam based on the current and/or past sensory data. Such methods result in frequent inference at each time step causing high overhead for sensing and processing in terms of power consumption, latency, and signaling. Long-term beam prediction can alleviate this by jointly predicting multiple future time steps, but this idea has remained largely unexplored. Existing work \cite{jiang2022computer,Jiang2024LiDAR} has considered long-term beam prediction based on vision and LiDAR data, respectively, achieving encouraging results. However, LiDAR is costly for widespread usage while a large model is required in \cite{jiang2022computer} to extract features from raw images.

To develop low-complexity and low-latency beam training, in this paper, we study efficient sensing-assisted long-term beam tracking designs by leveraging vision data to predict optimal beams for current and multiple future time slots. Specifically, given a predefined beamformer codebook, we formulate the beam tracking problem as an ML classification task, which is addressed using a sequence-to-sequence model. Distinct from \cite{jiang2022computer}, we adopt an efficient image preprocessing method followed by a dedicated convolutional neural network (CNN) to efficiently extract compact image feature. We further integrate an attention mechanism with gated recurrent unit (GRU) networks to capture the temporal dependencies of features across time slots, enabling robust long-term beam prediction. Numerical results on a real-world dataset demonstrate the superiority of the proposed design over the scheme in \cite{jiang2022computer} in terms of achieving comparable beam prediction accuracies across both current and six future time slots while reducing computational complexity by more than $97\%$. 

\section{System Model and Problem formulation}\label{sec:system model}
 \begin{figure*}[t]
\vspace{-10mm}
	\small
	\centering	
	\includegraphics[width=1\textwidth]{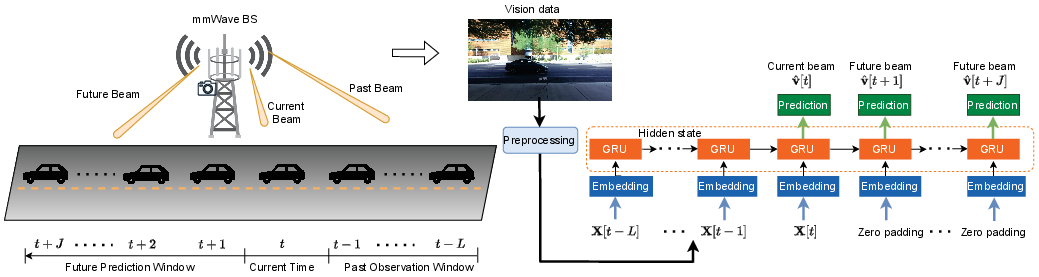}
	\vspace{-6mm}
	\caption{Illustration of the considered system model. The BS senses the environment and the moving UE with an RGB camera. The sensory data are collected and cached for beam tracking using the designed ML model. }
	\label{fig:system model}
	\vspace{-5mm}
\end{figure*}

{\bf System Model:}
We consider a downlink mmWave communications system, where the BS serves a single-antenna mobile user equipment (UE), as illustrated in Fig.~\ref{fig:system model}. The BS is equipped with a uniform linear array and an RGB camera (visual data sensor). At time step $t$, the BS transmits symbol $s[t]\in \Cs$, with $\Es\left[|s|^2\right]=1$, to the UE. We assume a block fading channel between time slots.  Let $\vb[t]$ denote the beamforming vector at time step $t$. Then, the received signal $y[t]$ can be written as
\begin{align}\label{eq:signal model}
y[t] = \hb[t]^\H \vb[t] s[t] + n[t],
\end{align}
where $\hb[t]$ denotes the channel between the BS and the UE at time step $t$, and $n[t]\sim\Ccl\Ncl(0, \sigma_{\rm n}^2)$ models the additive white Gaussian noise (AWGN) with noise power $\sigma_{\rm n}^2$. The signal-to-noise ratio (SNR) at time slot $t$ is $   \text{ SNR}[t]=\frac{|\hb[t]^\H \vb[t]\big|^2}{\sigma_{\rm n}^2}$.

{\bf Problem Formulation:}
At the current time slot $t$, our goal is to determine the transmit  beamforming vectors at the BS for current and $J$ future time slots, i.e., $\{t, t+1, \ldots, t+J\}$. 
Let $\Vcl = \{\vb_1,\ldots, \vb_{C}\}$ and $\Icl_{\Vcl}=\{1,\ldots,C\}$ denote the beamforming codebook and its associated index set with $C\triangleq|\Vcl|$. In the considered beam tracking problem, we aim to find $\vb[\tau] \in \Vcl, \forall \tau$, to maximize the overall spectral efficiency over these $J+1$ time slots, which is expressed as $R_J=\sum_{\tau=t}^{t+J}\log\left(1+ \text{ SNR}[\tau]\right)$. 
For low SNR scenarios, we can formulate the beamforming problem as \cite{Jiang2024LiDAR,imran2024environment,thomas2006elements}
\begin{equation}\label{pb:P2}
    \underset{\vb[\tau] \in \Vcl, \forall \tau}{\rm maximize} \quad  \sum_{\tau=t}^{t+J}|\hb[\tau]^\H \vb[\tau]\big|^2.
\end{equation}
Let $\bb^{\star}[t]=\big[b^{\star}[t],\ldots,b^{\star}[t+J]\big]^\T$ be the vector of beam indices corresponding to the optimal solution of \eqref{pb:P2}, i.e., 
\begin{equation}\label{pb:P3}
      \bb^{\star}[t] = \underset{b[\tau]\in \Icl_{\Vcl},\forall \tau}{\arg\max} \quad \sum_{\tau=t}^{t+J}|\hb[\tau]^\H \vb_{b[\tau]}\big|^2.
\end{equation}
The  solution to \eqref{pb:P3} can be obtained by decoupling it into $J+1$ subproblems with each solved via an exhaustive search over the $C$ candidate beams. However, the complexity of such a method scales as $J\cdot C$, which incurs high latency, especially with the large codebooks used in massive MIMO. Moreover, this approach requires perfect channel state information (CSI) at not only the current time slot, but also the \( J \) future time slots, which is generally unavailable in practice.

In this work, we consider CSI-free beam tracking, where instead of aiming to recover $\bb[t]$ based on knowledge of $\hb[t]$, we utilize sensed visual data, denoted by $\Zb[t]$. Accordingly, our aim is to design a CSI-free mapping from $\Zb[t]$ into $\bb[t]$, such that the results remains effective with respect to the CSI-based performance measure in \eqref{pb:P3}. Unlike \cite{Demirhan2022radar, Yang2023Environment,imran2024environment,charan2022vision,Xu20203D,cui2024sensing,Tariq2024deep,tian2023multimodal,shi2024multimodal, zhang2025multimodal,zhu2025advancing,park2025resource} which address problem \eqref{pb:P3} by decoupling it into $J+1$ subproblems, we propose an efficient learning framework that directly solves problem \eqref{pb:P3} for long-term beam tracking design, as will be elaborated below.




\section{Vision-based Long-term Beam Tracking}\label{sec:vision-based beam tracking}

{\bf Learning Task:} Let $\Zb[t]\in\Rs^{3\times d_{\rm H}\times d_{\rm W}}$ denote the RGB image obtained at time slot $t$, where the dimension $3$ corresponds to the number of RGB color channels, and $d_{\rm H}$ and $d_{\rm W}$ respectively represent the image height, and image width in pixels. Let $\Zcl[t] $ denote the sequence of sensory data, i.e., RGB images, from $L$ past time slots to the current time $t$, given by $\Zcl[t]=\{\Zb[t-L],\Zb[t-L+1], \ldots,\Zb[t]]\}$. We aim to develop an efficient ML model to solve \eqref{pb:P2}, i.e., to predict the optimal beams (equivalently the optimal beam indices in $ \Icl_{\Vcl}$) for current time slot $t$ and future $J$ time slots $t+1,\ldots, t+J$. Denote the data preprocessing operations by $\Xcl[t]=g(\Zcl[t])$, mapping the input sequence to the ML model. Let $f(\Xcl[t];\Theta)$ denote the ML model with learnable parameters $\Theta$. The ML model outputs the probabilities of all possible beams for beamforming at $J+1$ (current and future) time slots. Let $p_c[t + j]$ denote the probability of selecting the $c$-th beam in the codebook at time slot $t + j$, and define $\pb[t + j] = [p_1[t + j], \ldots, p_{C}[t + j]]^\T \in \Rs^{C}, j=0,\ldots,J$. 
The predicted beam index is obtained as
\begin{equation}\label{eq:predicted beam index}
    \hat{b}[\tau]=\arg\max_{c\in\Icl_{\Vcl}} \; p_c[\tau] ,\ \tau=t,\ldots,t+J.
\end{equation}
 The desired ML model for vision-aided beam tracking can be mathematically written as
\begin{equation}\label{pb:ML task}
    f^{\star}(;\Theta^{\star})=\arg\max_{f(;\Theta)} \quad \sum_{\tau=t}^{t+J} \Pbb\{\hat{b}[\tau]=b^{\star}[\tau]\},
\end{equation}
where $\Pbb\{\cdot\}$ denotes the probability. We note that $J$ and $L$ are hyperparameters, which are determined empirically.

 \begin{figure*}[t]
\vspace{-10mm}
	\small
	\centering	
	\includegraphics[width=1\textwidth]{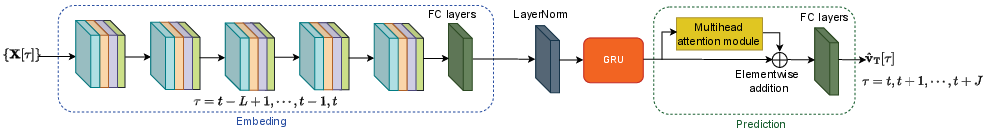}
	\vspace{-7mm}
	\caption{Illustration of the ML model structure. }
	\label{fig:ML model}
	\vspace{-6mm}
\end{figure*}

{\bf ML Model Structure:} As shown in Fig.~\ref{fig:system model}, the preprocessed input is firstly fed into the embedding block to extract semantic low-dimensional feature representation. The GRU block receives the feature vectors and captures their temporal dependencies by hidden states, which are updated step by step to produce a context vector that summarizes the information of the entire input $\{\Xb[\tau]\}$. Given this context vector and current data $\Xb[t]$, the current and future beams are predicted with the prediction block.
We adopt the standard GRU structure and focus on the design of the embedding and prediction blocks, which are illustrated in Fig.~\ref{fig:ML model}. A pretrained ResNet for image feature extraction \cite{cui2024sensing,Tariq2024deep,tian2023multimodal,shi2024multimodal,park2025resource} would result in a high cost of computational complexity and memory usage. To improve efficiency, we design a dedicated five-layer CNN network to extract spatial features of the images in the input sequence. Each layer consists of four standard Pytorch modules, including Cov2d, BatchNorm2d, ReLU, and MaxPool2d. The output of the GRUs is further processed by a residual multihead attention (MHA) module, which can strengthen the model's ability to capture temporal relations among the input sequence.

The MHA  aims to provide a self-attention mechanism that allows each feature in a sequence to attend to all others and simultaneously learn different contextual information components. For the mathematical foundations of MHA, we refer the reader to \cite{vaswani2017attention}. Using MHA self-attention after the GRU combines the strengths of both sequential and attention-based modeling. Specifically, while GRUs effectively capture temporal dependencies in a step-by-step manner, they focus mainly on local dependencies. In contrast, MHA self-attention allows the model to directly attend to all positions in the sequence, enhancing its ability to capture global features. Furthermore, multiple attention heads enable the extraction of diverse features from the GRU outputs, leading to richer and more informative representations that can improve the model’s expressive ability and boost performance. The effectiveness of the MHA module will be justified using numerical experiments in Section~\ref{sec:simulation}.


{\bf Training of the ML Model:}
Let $\Dcl_{\rm}=\{\{\Zcl[t], \bb^\star [t]\}, t=0,\ldots, T\}$ denote the set of data sequences from the source dataset, where $T$ denotes the number of time slots over which vision data are collected, $\Zcl[t]$ and $\bb^\star [t]$ are the ML model's input and label, respectively. The received signal $y[t]$ in \eqref{eq:signal model} is leveraged to obtain $\bb^\star [t]$, which in real-world mmWave communications is generally non-uniformly distributed among the $C$ candidate beams. Such class imbalance among the datasets can lead to poor performance for the minority class. During the training, we use the focal loss \cite{lin2017focal}, which is a modification of the standard cross-entropy loss designed to address the class imbalance problem.

 Given a sample $\Zb[\tau]$, the output of the ML model is the predicted probability vector $\pb [\tau]$ for the $C$ classes of beams. Let $\zb[\tau]=[z_1[\tau],\ldots,z_C[\tau]] $ denote the vector of the output logits. The $c$-th element of $\pb [\tau]$ is obtained as $p_c[\tau]=\sigma_c(\zb[\tau])$, where $\sigma_c(\zb[\tau])=\frac{\exp(z_c[\tau])}{\sum_{k=1}^C \exp(z_k[\tau]) }$ denotes the softmax function.  The focal loss for a single sample $\Zb[\tau]$ is given by
\begin{equation}\label{eq:Focal loss}
    l_{\rm focal}[\tau]=-(1-p_{b^{\star}}[\tau])^{\gamma}\log\left(p_{b^{\star}}[\tau] \right),
\end{equation}
where $p_{b^{\star}}[\tau]$ denotes the predicted probability of selecting the ground-truth beam index $b^{\star}[\tau]$ at time slot $\tau$. The hyperparameter $\gamma$ is the focusing parameter which down-weights easy examples. A large $\gamma$ leads to small loss for well-classified samples (those with high output probabilities), helping the model concentrate on difficult or misclassified samples that are more informative. On the contrary, $\gamma=0$ leads to the conventional cross-entropy loss which treats all samples with equal importance. The overall task loss for the input sequence $\Zcl[t]$ is expressed as
\begin{equation}\label{eq:task loss}
    \Lcl_{\rm task}[t]=\sum\nolimits_{\tau=t}^{t+J} l_{\rm focal}[\tau].
\end{equation}
The overall training procedure is summarized in Algorithm~~\ref{alg1}, where $E$, $N_{\rm b}$, and $B$ represents the number of total epochs, the number of batches in each epoch,  and batch size, respectively. Here, $\Dcl_{\rm tr}$ and $\Dcl_{\rm evl}$ represents the training and validation datasets, respectively. With random initialization, the model is updated over $E$ epochs. For each epoch, $N_{\rm b}$ batches are randomly generated for batch training in steps 4--9, where $\Tcl^{(n)}=\{t_i^{(n)} \big| \Zcl[t_i^{(n)}] \in \Dcl^{(n)}_{\rm tr}, \forall i\}=\{t_1^{(n)},\ldots,t_B^{(n)}\}$ denotes the set of time stamps in the $n$-th batch. The best model is updated if a lower validation loss is found after batch training.

\begin{algorithm}[t]
\small
\caption{Learning Procedure for Problem~\eqref{pb:ML task}.}\label{alg1}
\LinesNumbered 
\KwIn{Training and validation datasets $\Dcl_{\rm tr}$, $\Dcl_{\rm evl}$}
\KwOut{Model parameters $\Theta$}
 Initialize $\Theta^{\star}$, $\Theta=\Theta^{\star}$, and learning rate.\\
\For{$e=1,\ldots,E$}{
    Randomly divide $\Dcl_{\rm tr}$ into $N_{\rm b}$ batches $\{\Dcl^{(n)}_{\rm tr}\}_{n=1}^{N_{\rm b}}$ with batch size $B$. \\
    \For{$n=1,\ldots,N_{\rm b}$}{
        Perform data preprocessing $\Xcl[t]=g\left(\Zcl[t]\right)$ with $t\in \Tcl^{(n)}=\{t_1^{(n)},\ldots,t_B^{(n)}\}$ \\

        Feed the sequences $\Xcl[t_q^{(n)}],q=1,\ldots,B$ into the the ML model and compute $\pb_q[t+j],j=0,1,\ldots,J,q=1,\ldots, B$. \\
        

                 
        Compute average task loss over the batch in $J+1$ time slots: $\Lcl_{\rm task}=\frac{1}{B(J+1)}\sum\nolimits_{ t\in \Tcl^{(n)}} \Lcl_{\rm task}[t]$.\\
        Update $\Theta$ with an optimizer\\

    }
    Compute validation loss based on $f(;\Theta)$ and $\Dcl_{\rm evl}$.\\
    Update the best model $\Theta^{\star}=\Theta$ if a lower validation loss is found.
 }

    Return $\Theta^{\star}$.
\end{algorithm}

\vspace{-3mm}
\section{Numerical Results}\label{sec:simulation}
 In this section, we evaluate the performance of the proposed ML model for beam tracking.\footnote{The  code source is available online at \url{https://github.com/WillysMa/Sensing-Assisted-Beam-Tracking.git}.} Experiments are based on Scenario 9 of the DeepSense 6G dataset \cite{alkhateeb2023deepsense}, which provides sensory data and optimal beams for real-world mmWave communications. For any time step $t$, the maximum number of past images is set to $L=8$ in each sequence sample $\Zcl[t]$, while the number of future time steps for beam prediction is set to $J=6$. The overall dataset contains a total of $T=4060$ samples $\{\Zcl[t], \bb^{\star}[t]\}$ with $80\%$ training samples and $20\%$ validation samples. 
 The ML models are implemented by PyTorch and trained via NViDIA Tesla V100 GPUs. In the training stage, the initial learning rate is $10^{-4}$ with a cyclic cosine annealing scheduler used. We set $\gamma=2$ for the loss function. In the MHA module, we adopt $8$ heads for diverse attentions. As a result, the ML model has approximately a total of $1.8 \times 10^6$ trainable parameters. Further details on the experimental setup are available in the released source code.

\begin{table}[t]
\small
\centering
\caption{Overall generalization performance of the ML model in $\%$ for ATop-$k$ accuracy and ADBA score.}\label{tb:teacher model}
\vspace{-2mm}
\begin{tabular}{r|c|c|c}
\hline
Metric &W/o MHA &   With MHA         & Optimal \cite{jiang2022computer} \\
\hline
Test loss & $1.141$    &   $1.050$               &     $0.8158$  \\
\hline
 ATop-$1$ & $40.77$  &     $42.91$             &     $50.20$        \\
 \hline
 ATop-$3$ & $77.44$  &     $79.97$              &     $ 87.15$        \\
 \hline
  ATop-$5$ & $92.31$  &     $93.35$              &     $96.81$        \\

\hline
 ADBA & $93.50$  &     $94.47$                 &     $96.63$        \\

\hline
\end{tabular}
\end{table}

We use the task loss, Top-$k$ accuracy, and distance-based accuracy (DBA) \cite{charan2022multi} for performance evaluation. The task loss at the inference stage reflects the overall generalization performance, while the other two metrics targets for specific time slots. Specifically, the Top-$k$ accuracy measures whether the ground-truth label is among the model's Top-$k$ predicted labels. 
In contrast, the DBA metric computes the distance of the predicted beams from the ground-truth beams and assigns scores based on how far the predicted beam is from the ground-truth beam. The DBA score is computed using Top-$3$ accuracy, and we refer readers to \cite{charan2022multi} for a detailed explanation. 
 Note that both the Top-$k$ and DBA scores target one time slot. To reflect the overall performance across all $J+1$ time slots, we further consider average Top-$k$ (ATop-$k$) and average DBA (ADBA) over all time slots.

\begin{figure}[t]
\small
    \centering
    \hspace{-5mm}
    \subfigure[Top-$1$ accuracy.]
    {\label{fig:top1}\includegraphics[width=0.24\textwidth]{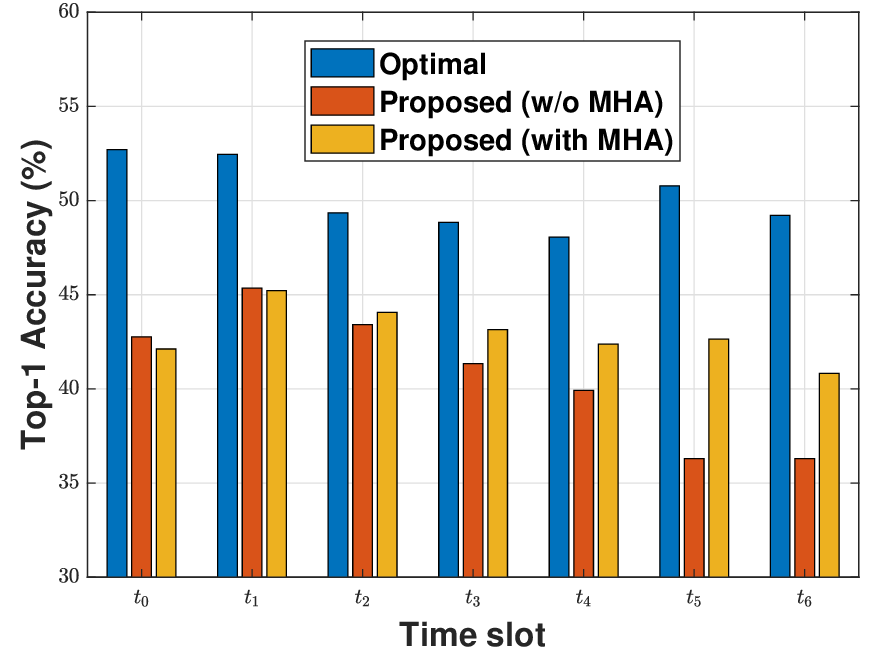}}
     \hspace{-2mm}
    \subfigure[Top-$3$ accuracy.]
    {\label{fig:top3} \includegraphics[width=0.24\textwidth]{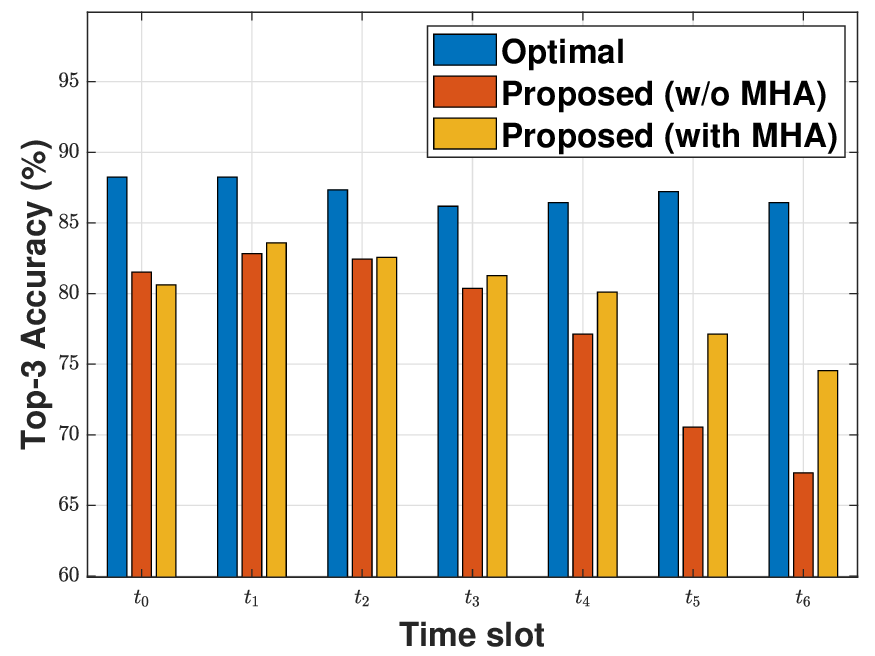}}
        \hspace{-5mm}
    \subfigure[Top-$5$ accuracy.]
    {\label{fig:top5}\includegraphics[width=0.24\textwidth]{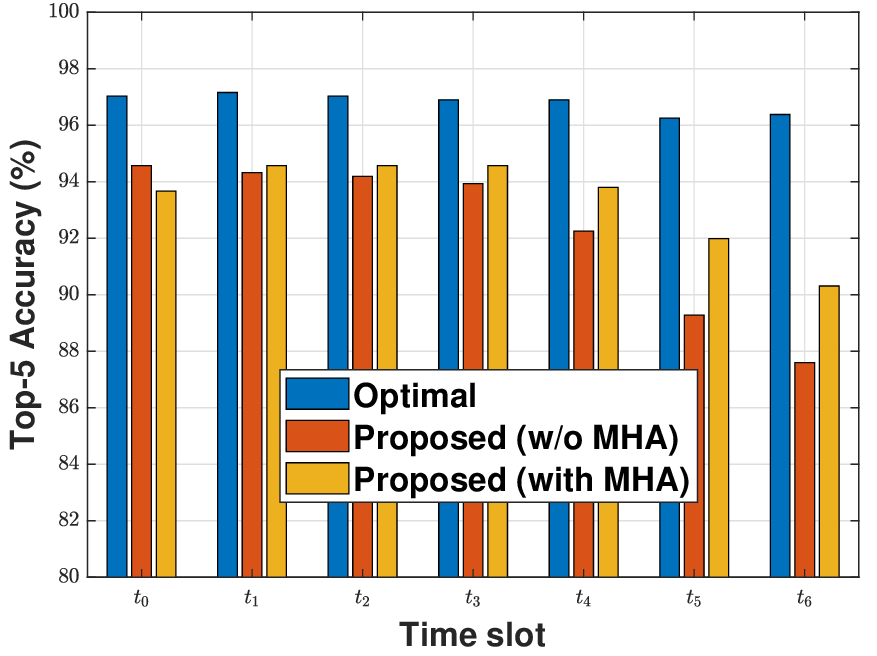}}
     \hspace{-2mm}
    \subfigure[DBA score.]
    {\label{fig:dba} \includegraphics[width=0.24\textwidth]{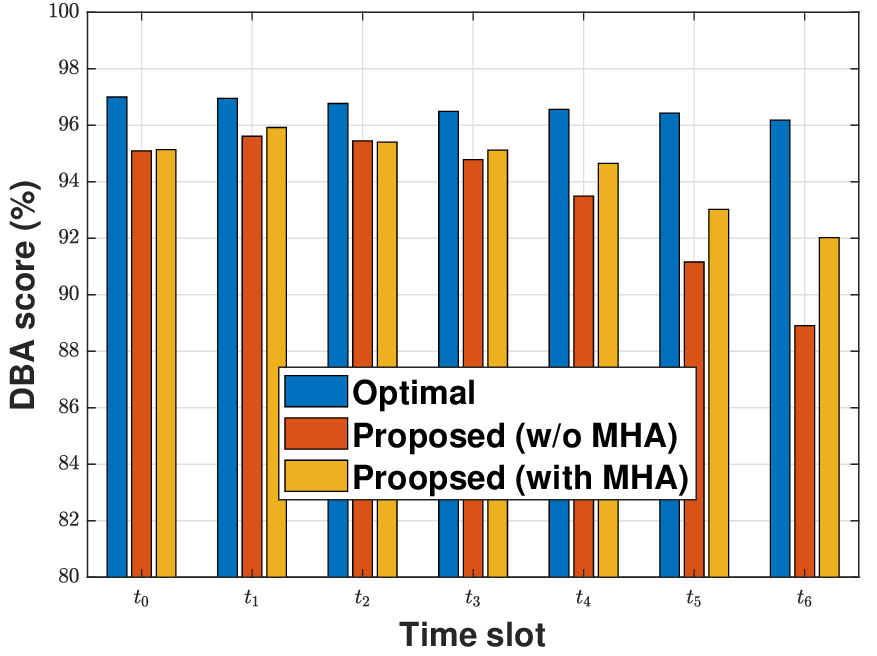}}
    \vspace{-3mm}
    \caption{Performance of the ML model.}
    \label{fig:performance}
\end{figure}

 Table~\ref{tb:teacher model} shows the overall generalization performance of the designed ML model, where ``W/o MHA'' and ``With MHA'' represent the architecture without and with the MHA module. Here, the testing loss equals the validation loss because the same dataset is used for both. 
 Unlike \cite{jiang2022computer}, which relies on YOLOv4 \cite{bochkovskiy2020yolov4}  and manual selection to eliminate the interference to the sensing target, we suppress background interference through adjacent-frame subtraction and highlight the sensing target with motion masks. Together with the designed CNNs, we extract compact image features at significantly lower complexity than YOLOv4 (approximately $6.4\times 10^7$ parameters). It is seen that the MHA module can effectively enhance the generalization of the ML model. Furthermore, the proposed ML model (with MHA) can achieve over $97\%$ of the optimal ADBA with only $3\%$ complexity of the model in \cite{jiang2022computer} (including the YOLOv4). We note that the model complexity is evaluated in terms of the total number of trainable parameters.
 
 Figs.~\ref{fig:top1}--\ref{fig:dba} show the Top-$1$, Top-$3$, and Top-$5$ accuracy and DBA score of the considered schemes versus current and six future time slots, respectively. All schemes show lower accuracy and DBA scores for distant time slots, as beam prediction becomes more difficult further into the future. Furthermore, we observe that the integrated attention mechanism primarily improves the prediction accuracy of far future time slots, i.e., $t_4$--$t_6$, especially for Top-$1$ and Top-$3$ accuracy. For instance, percentage-point gains of $6.34$ in Top-$1$ accuracy, $6.59$ in Top-$3$ accuracy, $2.7$ in Top-$5$ accuracy, and $1.86$ in DBA score at time slot $t_5$ are attained by the attention-enhanced ML model compared to the architecture without MHA. Such results verify that the MHA can effectively capture the time dependency relation of the input sequence, extracting useful information for more challenging future beam predictions. Although the Top-$3$ accuracy of the proposed ML model is no more than $85\%$, the corresponding DBA scores can reach up to $95\%$, as seen in Fig.~\ref{fig:dba}.

\vspace{-3mm}
\section{Conclusions}
This paper proposed an attention-enhanced end-to-end learning framework for long-term beam tracking based on past sensing images. We designed an efficient ML model consisting of CNNs, GRUs, and MHA for extracting features and processing temporal dependencies. Numerical results demonstrate the superiority of the proposed ML model over the state-of-the-art scheme in terms of achieving comparable beam prediction accuracies across both current and six future time slots while requiring only $3\%$ its complexity. The designed long-term beam tracking significantly reduces signaling overhead, latency, and power consumption associated with sensing and processing for beam training, advancing the practical implementation. Generalization under domain shifts and robustness in multi-user scenarios are currently constrained by the low-complexity design and dataset limitations, and are left for future study.

\clearpage
    \vspace*{-3cm}
\let\OLDthebibliography\thebibliography
\renewcommand\thebibliography[1]{
\vspace*{-4mm}
    \OLDthebibliography{#1}
    \setlength{\parskip}{0pt}
    \setlength{\itemsep}{0pt plus 0.0ex}
    \vspace{-3mm}
}

\section{Acknowledgments}
This work was supported by the Research Council of Finland through 6G Flagship Program (grant 369116) and projects DIRECTION (grant 354901), DYNAMICS (grant 24305016), and CHIST-ERA PASSIONATE (grant 359817).

\bibliographystyle{IEEEbib}
\bibliography{IEEEabrv,reference}

@STRING{IEEE_J_STSP       = "{IEEE} J. Sel. Topics Signal Process."}

@STRING{IEEE_J_JSAC       = "{IEEE} J. Sel. Areas Commun."}

@STRING{IEEE_J_COM        = "{IEEE} Trans. Commun."}

@STRING{IEEE_J_WCOM       = "{IEEE} Trans. Wireless Commun."}

@STRING{IEEE_J_WCOML      = "{IEEE} Wireless Commun. Lett."}

@STRING{IEEE_J_IOT        = "{IEEE} Internet Things J."}

@STRING{IEEE_M_COM        = "{IEEE} Commun. Mag."}

@string{ globecomw = {Proc. IEEE Global Commun. Conf. Workshop}}

@string{ icc = {Proc. IEEE Int. Conf. Commun.}}

@string{ vtc = {Proc. IEEE Veh. Technol. Conf.}}

@string{ wcnc = {Proc. IEEE Wireless Commun. and Networking Conf.}}

@string{ eusipco = {Proc. European Sign. Proc. Conf.}}

@ARTICLE{Jiang2024LiDAR,
  author={Jiang, Shuaifeng and Charan, Gouranga and Alkhateeb, Ahmed},
  journal=IEEE_J_WCOML,
  title={{LiDAR} Aided Future Beam Prediction in Real-World Millimeter Wave {V2I} Communications},
  year={2023},
  volume={12},
  number={2},
  pages={212-216},
  doi={10.1109/LWC.2022.3219409}}

@INPROCEEDINGS{Demirhan2022Radar,
  author={Demirhan, Umut and Alkhateeb, Ahmed},
  booktitle=wcnc,
  title={Radar Aided {6G} Beam Prediction: Deep Learning Algorithms and Real-World Demonstration},
  year={2022},
  volume={},
  number={},
}

@inproceedings{charan2022vision,
  title={Vision-position multi-modal beam prediction using real millimeter wave datasets},
  author={Charan, Gouranga and Osman, Tawfik and Hredzak, Andrew and Thawdar, Ngwe and Alkhateeb, Ahmed},
  booktitle=wcnc,
  year={2022}
}

@inproceedings{morais2023position,
  title={Position-aided beam prediction in the real world: How useful {GPS} locations actually are?},
  author={Morais, Jo{\~a}o and Bchboodi, Arash and Pezeshki, Hamed and Alkhateeb, Ahmed},
  booktitle=icc,
  year={2023}
}

@article{alkhateeb2023deepsense,
  title={DeepSense {6G}: A large-scale real-world multi-modal sensing and communication dataset},
  author={Alkhateeb, Ahmed and Charan, Gouranga and Osman, Tawfik and Hredzak, Andrew and Morais, Joao and Demirhan, Umut and Srinivas, Nikhil},
  journal=IEEE_M_COM,
  volume={61},
  number={9},
  pages={122--128},
  year={2023},
  publisher={IEEE}
}

@article{charan2022multi,
  title={Multi-modal beam prediction challenge 2022: Towards generalization},
  author={Charan, Gouranga and Demirhan, Umut and Morais, Jo{\~a}o and Behboodi, Arash and Pezeshki, Hamed and Alkhateeb, Ahmed},
  journal={arXiv preprint arXiv:2209.07519},
  year={2022}
}

@inproceedings{jiang2022computer,
  title={Computer vision aided beam tracking in a real-world millimeter wave deployment},
  author={Jiang, Shuaifeng and Alkhateeb, Ahmed},
  booktitle=globecomw,
  year={2022}
}

@article{bochkovskiy2020yolov4,
  title={{Yolov4}: Optimal speed and accuracy of object detection},
  author={Bochkovskiy, Alexey and Wang, Chien-Yao and Liao, Hong-Yuan Mark},
  journal={arXiv preprint arXiv:2004.10934},
  year={2020}
}

@article{imran2024environment,
  title={Environment semantic communication: Enabling distributed sensing aided networks},
  author={Imran, Shoaib and Charan, Gouranga and Alkhateeb, Ahmed},
 journal=IEEE_J_OJCOMS,
  year={2024},
  volume={5},
  number={},
  pages={7767-7786},
  publisher={IEEE}
}

@article{vaswani2017attention,
  title={Attention is all you need},
  author={Vaswani, Ashish and Shazeer, Noam and Parmar, Niki and Uszkoreit, Jakob and Jones, Llion and Gomez, Aidan N and Kaiser, {\L}ukasz and Polosukhin, Illia},
  journal={Advances in neural information processing systems},
  volume={30},
  year={2017}
}

@article{park2025resource,
  title={Resource-Efficient Beam Prediction in mmWave Communications with Multimodal Realistic Simulation Framework},
  author={Park, Yu Min and Tun, Yan Kyaw and Saad, Walid and Hong, Choong Seon},
  journal={arXiv preprint arXiv:2504.05187},
  year={2025}
}

@article{zhu2025advancing,
  title={Advancing multi-modal beam prediction with cross-modal feature enhancement and dynamic fusion mechanism},
  author={Zhu, Qihao and Wang, Yu and Li, Wenmei and Huang, Hao and Gui, Guan},
  journal=IEEE_J_COM,
  year={Early access, 2025},
  publisher={IEEE}
}

@article{cui2024sensing,
  title={Sensing-assisted high reliable communication: A transformer-based beamforming approach},
  author={Cui, Yuanhao and Nie, Jiali and Cao, Xiaowen and Yu, Tiankuo and Zou, Jiaqi and Mu, Junsheng and Jing, Xiaojun},
  journal=IEEE_J_STSP,
  volume={18},
  number={5},
  pages={782--795},
  year={2024},
  publisher={IEEE}
}

@inproceedings{lin2017focal,
  title={Focal loss for dense object detection},
  author={Lin, Tsung-Yi and Goyal, Priya and Girshick, Ross and He, Kaiming and Doll{\'a}r, Piotr},
  booktitle={Proceedings of the IEEE international conference on computer vision},
  pages={2980--2988},
  year={2017}
}

@ARTICLE{Rezaie2022deep,
  author={Rezaie, Sajad and de Carvalho, Elisabeth and Manchón, Carles Navarro},
  journal=IEEE_J_WCOM,
  title={A Deep Learning Approach to Location- and Orientation-Aided {3D} Beam Selection for {mmWave} Communications},
  year={2022},
  volume={21},
  number={12},
  pages={11110-11124}
}

@ARTICLE{Xu20203D,
  author={Xu, Weihua and Gao, Feifei and Jin, Shi and Alkhateeb, Ahmed},
  journal=IEEE_J_WCOML,
  title={{3D} Scene-Based Beam Selection for {mmWave} Communications},
  year={2020},
  volume={9},
  number={11},
  pages={1850-1854}
}

@INPROCEEDINGS{Luo2023millimeter,
  author={Luo, Hao and Demirhan, Umut and Alkhateeb, Ahmed},
  booktitle=eusipco,
  title={Millimeter Wave {V2V} Beam Tracking using Radar: Algorithms and Real-World Demonstration},
  year={2023}
}

@ARTICLE{Klautau2019LIDAR,
  author={Klautau, Aldebaro and González-Prelcic, Nuria and Heath, Robert W.},
  journal=IEEE_J_WCOML,
  title={{LIDAR} Data for Deep Learning-Based {mmWave} Beam-Selection},
  year={2019},
  volume={8},
  number={3},
  pages={909-912}
}

@ARTICLE{Yang2023Environment,
  author={Yang, Yuwen and Gao, Feifei and Tao, Xiaoming and Liu, Guangyi and Pan, Chengkang},
  journal=IEEE_J_JSAC,
  title={Environment Semantics Aided Wireless Communications: A Case Study of {mmWave} Beam Prediction and Blockage Prediction},
  year={2023},
  volume={41},
  number={7},
  pages={2025-2040}
}

@ARTICLE{Tariq2024deep,
  author={Tariq, Shehbaz and Arfeto, Brian Estadimas and Khalid, Uman and Kim, Sunghwan and Duong, Trung Q. and Shin, Hyundong},
  journal=IEEE_J_IOT,
  title={Deep Quantum-Transformer Networks for Multimodal Beam Prediction in {ISAC} Systems},
  year={2024},
  volume={11},
  number={18},
  pages={29387-29401}
}

@article{tian2023multimodal,
  title={Multimodal transformers for wireless communications: A case study in beam prediction},
  author={Tian, Yu and Zhao, Qiyang and Boukhalfa, Fouzi and Wu, Kebin and Bader, Faouzi and others},
  journal={arXiv preprint arXiv:2309.11811},
  year={2023}
}

@inproceedings{shi2024multimodal,
  title={Multimodal deep learning empowered millimeter-wave beam prediction},
  author={Shi, Binpu and Li, Min and Zhao, Ming-Min and Lei, Ming and Li, Liyan},
  booktitle=vtc,
  year={2024}
}

@article{zhang2025multimodal,
  title={Multimodal deep learning-empowered beam prediction in future {THz ISAC} systems},
  author={Zhang, Kai and Yu, Wentao and He, Hengtao and Song, Shenghui and Zhang, Jun and Letaief, Khaled B},
  journal={arXiv preprint arXiv:2505.02381},
  year={2025}
}

@book{thomas2006elements,
  title={Elements of information theory},
  author={Thomas, MTCAJ and Joy, A Thomas},
  year={2006},
  publisher={Wiley-Interscience}
}

\end{document}